\input{aipcheck}

\documentclass[
    ,final            
  ]
  {aipproc}

\layoutstyle{8x11double}
\usepackage{bm}
\usepackage{amsmath}
\usepackage[abs]{overpic}

\begin{document}

\title{Pattern dynamics of cohesive granular particles\\under a plane shear}


\author{Satoshi Takada}{}

\author{Hisao Hayakawa}{
  address={Yukawa Institute for Theoretical Physics, Kyoto University, Kitashirakawa Oiwakecho, Sakyo-ku, Kyoto 606-8502 Japan}
}

\begin{abstract}
We perform three dimensional molecular dynamics simulations of cohesive granular particles under a plane shear.
From the simulations, we found that the granular temperature of the system abruptly decreases to zero
after reaching the critical temperature, where the characteristic time $t_{\rm cl}$ is approximately represented by
$t_{\rm cl} \propto (\zeta-\zeta_{\rm cr})^{-\beta}$ with the dissipation rate $\zeta$
, the critical dissipation rate $\zeta_{\rm cr}$ and the exponent $\beta \simeq 0.8$.
We also found that there exist a variety types of clusters depending on the initial density and the dissipation rate.
\end{abstract}

\maketitle


\section{Introduction}
\quad The interactions among macroscopic granular particles,
such as sands and granules, are characterized by a repulsive and a dissipative ones.
The energy dissipation through inelastic collisions causes destabilization of a uniform state of the system.
It is known that there appear clusters in such systems \cite{Goldhirsch, Goldhirsch2},
which may be understood by the hydrodynamic equations \cite{McNamara, McNamara2, Brilliantov, Brilliantov2}.
When a shear is applied to the granular system,
there appear high-density region, called "shear band"
in the two dimensional systems \cite{Saitoh2007, Conway},
and various shapes of clusters such as 2D plug, 2D wave or 3D wave
for three dimensional systems \cite{Conway, Lasinski}.
There exist many papers to estimate the transport coefficients by using
kinetic theory
\cite{Jenkins1983,Lutsko2005,Garzo1999,Lun1984,Sela1996,
Montanero1999} and to analyze the pattern dynamics by using continuum mechanics
\cite{Saitoh2007,Saitoh2011,Alam2012,Alam1998,Alam1997,Savage1992,Garzo2006,Schmid1994,Gayen2006,Nott1999}.

On the other hand, fine powders of submicron order, such as tonner particles or interstellar dusts,
have attractive forces like an electrostatic force \cite{Castellanos}.
The existence of the attractive force causes some new features due to competition of 
the gas-liquid phase transition \cite{Hansen, Smit} and the dissipative structure \cite{Saitoh2007}.
For instance, nucleation process near equilibrium is well understood \cite{Yasuoka},
but that under a shear has not been well understood yet.
In this paper, we try to characterize the nonequilibrium pattern formation of fine powders 
under a plane shear based on the three dimensional molecular dynamics simulation.

In our previous paper \cite{ST}, we have fixed the dissipation rate and
clarified how the patterns depend on the density and the shear rate.
In this paper, on the other hand, we mainly focus on the role of the dissipation rate to the pattern dynamics.

\section{Model and Setup}
\quad The system we consider consists of monodiperse $N(=10,000)$ spheres,
whose radius is $\sigma$ and mass is $m$.
The system is a cubic and its linear size is $L$.
We choose $x$-axis and $y$-axis as the direction of the shear velocity and that of the velocity gradient, respectively.
The interaction among particles is assumed to be described by the truncated Lennard-Jones potential
\begin{equation}
U^{\rm LJ}(r_{ij})=
\begin{cases}
4\varepsilon\left[ \left( \frac{\sigma}{r_{ij}} \right)^{12} - \left(
\frac{\sigma}{r_{ij}} \right)^{6} \right] & (r\leq r_{\rm c})\\
0 & (r > r_{\rm c})
\end{cases},
\end{equation}
with the well depth $\varepsilon$, where $r_{ij}=|\bm{r}_{ij}|$ is the distance between $i$-th and $j$-th 
($1\leq i , j \leq N$) particles
and $r_{\rm c}$ is cut-off length (in this paper, we use $r_{\rm c}=3\sigma$).
For the dissipation force, we use
\begin{eqnarray}
\bm{F}^{\rm vis}(\bm{r}_{ij},\bm{v}_{ij}) = -\zeta \Theta (\sigma - |\bm{r}_{ij}|) (\bm{v}_{ij} \cdot \hat{\bm{r}}_{ij})\hat{\bm{r}}_{ij},
\end{eqnarray}
with the dissipation rate $\zeta$, where $\bm{v}_{ij}$ is the relative velocity vector of $i$-th and $j$-th particles
defined by $\bm{v}_{ij}=\dot{\bm{r}}_{ij}$,
$\Theta(r)$ is the step function which is one for $r>0$ and zero for otherwise,
and $\hat{\bm{r}}$ is a unit vector proportional to $\bm{r}$.
We note that $\zeta$ is a dissipation parameter related to the coefficient of restitution $e$,
for example, $e = 0.998$ for $\zeta = 0.1$ and $e= 0.983$ for $\zeta = 1.0$, respectively, at the temperature $T=1.4\varepsilon$.
Thus we are interested in weakly dissipative situations.
This weak dissipation is necessary to reach a steady state.
Finally, the force acting on $i$-th particle is given by
\begin{equation}
\bm{F}_i = -\sum_{j \neq i} \bm{\nabla}_i U^{\rm LJ}(r_{ij}) + \sum_{j \neq i} \bm{F}^{\rm vis}(\bm{r}_{ij},\bm{v}_{ij}).
\end{equation}

In general, we cannot ignore the boundary effect \cite{Karison, Richman, Dave, Campbell, Tan}.
In order to suppress the boundary effect to the system, 
we adopt the Lees-Edwards periodic boundary condition \cite{LEbc, EvansLiquids}
with the aid of SLLOD algorithm \cite{EvansLiquids, EvansThesis}:
\begin{eqnarray}
\frac{d\bm{r}_i}{dt} &=& \frac{\bm{p}_i}{m} + \dot\gamma y_i \hat{\bm{e}}_x,\\
\frac{d\bm{p}_i}{dt} &=& \bm{F}_i - \dot\gamma p_{yi} \hat{\bm{e}}_x.
\end{eqnarray}
Here $\bm{r}_i=(x_i,y_i,z_i)$, $\bm{p}_i=(p_{xi},p_{yi},p_{zi})$ are the position and the momentum 
of $i$-th particle, respectively,
$\dot\gamma$ is the shear rate and $\hat{\bm{e}}_x$ is the unit vector in $x$ direction.
This algorithm is known that the system is relaxed to a uniform shear state near equilibrium.
We equilibrate the system without the shear and the dissipation
until $t=200(m\sigma^2/\varepsilon)^{1/2}$,
and then we apply the shear and the dissipation to the system.

\section{Results}
\begin{figure}
	\begin{minipage}{0.49\hsize}
	\begin{overpic}[width=38mm]{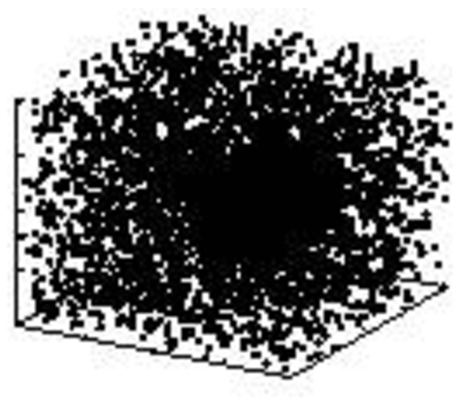}
		\put(31,-3){$x^\ast$}
		\put(-8,38){$y^\ast$}
		\put(97,5){$z^\ast$}
		\put(55,-5){(a)}
	\end{overpic}
	\end{minipage}
	\begin{minipage}{0.49\hsize}
	\begin{overpic}[width=38mm]{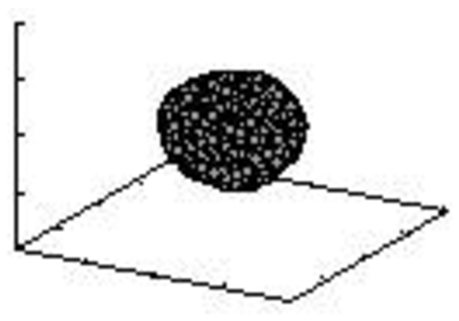}
		\put(31,-3){$x^\ast$}
		\put(-5,38){$y^\ast$}
		\put(97,5){$z^\ast$}
		\put(55,-13){(b)}
	\end{overpic}
	\end{minipage}
	\caption{Typical time evolution of the system for the dilute case $\bar\rho=0.0463$
				(${\dot\gamma}^\ast=0.01$, $\zeta^\ast=0.0173$), 
				(a) transitional regime at $t^\ast=65,500$ and (b) steady regime at $t^\ast=92,000$.}
	\label{fig:evol}
\end{figure}

\quad At first, we consider a dilute case $\bar\rho=0.0463$,
where $\bar\rho$ is the average density $N\sigma^3/L^3$.
The system size is $L=60\sigma$ for this density.
We set $T_0 = 1.0\varepsilon$ as the initial temperature, 
which is slightly higher than the critical temperature
for the equilibrium Lennard-Jones fluid \cite{Hansen, Smit}.
We fix the shear rate $\dot\gamma$
and change the average density $\bar\rho$ and the dissipation rate $\zeta$.
From our simulation, we found two steady phases: (i) uniformly sheared phase and (ii) clustered phase.
In this paper, we focus on the (ii) phase.
In this phase, the shape of clusters is a spherical droplet for low density.
A typical time evolution of the configuration of the system is drawn in Fig. \ref{fig:evol},
where $\bm{r}^\ast = \bm{r}/\sigma$, ${\dot\gamma}^\ast = \dot\gamma (m\sigma^2/\varepsilon)^{1/2}$,
$\zeta^\ast =\zeta (m\sigma^2/\varepsilon)^{1/2}$ and $t^\ast = t(\varepsilon/m\sigma^2)^{1/2}$.
There appears one big droplet, and all particles in the system are finally absorbed in it.
This result corresponds to the case for small shear rate in our previous paper \cite{ST}.

\begin{figure}
	\begin{overpic}{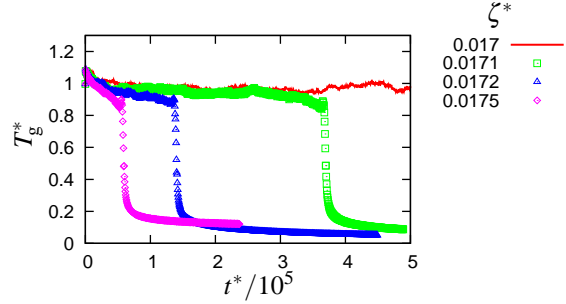}
	\put(80,-5){$t^\ast/10^5$}
	\put(0,50){\rotatebox{90}{$T_{\rm g}^\ast$}}
	\put(179,98){$\zeta^\ast$}
	\end{overpic}
	\caption{Typical time evolution of the granular temperature
				$T_{\rm g}$ for $\bar\rho=0.0463$, ${\dot\gamma}^\ast=0.01$
				with several $\zeta^\ast$.}
	\label{fig:temperature}
\end{figure}

We try to characterize this nucleation process and the droplet growth
 in terms of a simple physical argument based on the granular temperature 
$T_{\rm g} = (m/3N)\sum_{i=1}^N |\bm{v}_i-\bar{\bm{v}}|^2$ \cite{Tg1, Tg2},
where $\bar{\bm{v}} = \bar{\bm{v}}(\bm{r},t)$ is the average velocity as a function of $\bm{r}$.
The time evolution of $T_{\rm g}$ is plotted in Fig. \ref{fig:temperature},
where $T_{\rm g}^\ast =T_{\rm g}/\varepsilon$.
From this figure, 
we found that there exist a critical temperature $T_{\rm cr}$
and the the critical dissipation rate $\zeta$ corresponding to $T_{\rm cr}$.
In this density, $T_{\rm cr}$ is approximately equal to $0.9\varepsilon$.
Here we note that this temperature is slightly lower than the equilibrium value \cite{Hansen, Smit}.
For $\zeta < \zeta_{\rm cr}$,
the energy gain by the shear and the energy loss by the dissipation are balanced,
which keeps the system uniform.
On the other hand, for $\zeta > \zeta_{\rm cr}$,
at an early stage, the system remains uniform and 
$T_{\rm g}$ decreases gradually from the initial temperature $T_0$ to the critical temperature $T_{\rm cr}$ as time goes on.
Then, after reaching $T_{\rm cr}$,
there appears a big droplet (Fig. \ref{fig:evol}(a)) and it absorbs all particles in the system.
After reaching $T_{\rm g} = T_{\rm cr}$, $T_{\rm g}$ abruptly decreases to become zero.
The time when $T_{\rm g}$ begins to decrease corresponds to the time at which the nucleation takes place.

\begin{figure}
	\begin{overpic}[width=75mm]{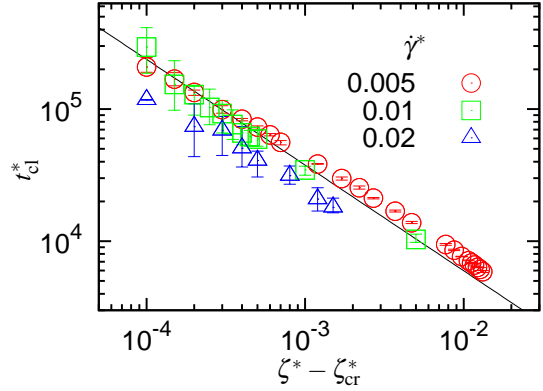}
	\put(95,-5){$\zeta^\ast-\zeta_{\rm cr}^\ast$}
	\put(-5,70){\rotatebox{90}{$t_{\rm cl}^\ast$}}
	\put(143,118){${\dot\gamma}^\ast$}
	\end{overpic}
	\caption{The dependence of the clustering time $t_{\rm cl}^\ast$ on the dissipation rate 
				$\zeta^\ast-\zeta_{\rm cr}^\ast$ for $\bar\rho=0.0463$,
				where the solid line denotes $t_{\rm cl} \propto (\zeta-\zeta_{\rm cr})^{-4/5}$.}
	\label{fig:clustering}
\end{figure}

In order to characterize the nucleation process
we introduce the ``clustering time'' $t_{\rm cl}$, which is defined by the time 
when $T_{\rm g}$ begins to drop as in Fig. \ref{fig:temperature}.
The dependence of $t_{\rm cl}$ on $\zeta$ is plotted in Fig. \ref{fig:clustering},
where $t_{\rm cl}^\ast = t_{\rm cl} (\varepsilon/m\sigma^2)^{1/2}$ and 
$\zeta_{\rm cr}^\ast=\zeta_{\rm cr} (m\sigma^2/\varepsilon)^{1/2}$.
This is approximately given by
\begin{eqnarray}
t_{\rm cl} = \alpha (\zeta- \zeta_{\rm cr})^{-\beta} \quad (\zeta > \zeta_{\rm cr}),
\end{eqnarray}
where $\alpha$ and $\beta$ are constants and $\zeta_{\rm cr}$ depends on the shear rate.
Here we note that $\zeta_{\rm cr}$ is a fitting parameter, and 
we evaluate $\beta\simeq 0.8$ and $\zeta_{\rm cr}^\ast = 0.0043$, $0.017$ and $0.068$
for ${\dot\gamma}^\ast=0.005$, $0.01$ and $0.02$, respectively.

\begin{figure}
	\begin{overpic}[width=79.7mm]{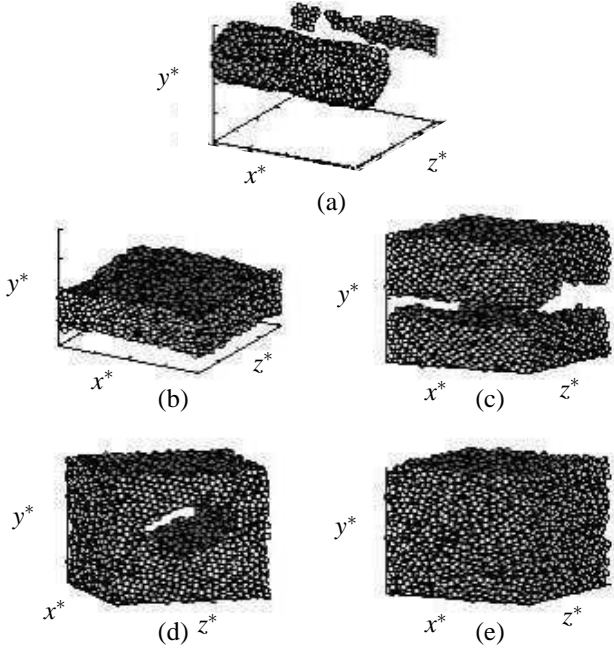}
		\put(83,165){$x^\ast$}
		\put(50,202){$y^\ast$}
		\put(152,170){$z^\ast$}
		\put(110,155){(a)}
		\put(25,87){$x^\ast$}
		\put(-7,124){$y^\ast$}
		\put(86,94){$z^\ast$}
		\put(50,80){(b)}
		\put(151,83){$x^\ast$}
		\put(118,120){$y^\ast$}
		\put(202,84){$z^\ast$}
		\put(170,80){(c)}
		\put(7,0){$x^\ast$}
		\put(-5,36){$y^\ast$}
		\put(65,-5){$z^\ast$}
		\put(50,-8){(d)}
		\put(151,-5){$x^\ast$}
		\put(117,32){$y^\ast$}
		\put(203,-2){$z^\ast$}
		\put(170,-8){(e)}
	\end{overpic}
	\caption{Typical shapes of clusters: (a) 2D plug, (b) 2D plate, (c) 2D inverse plate,
				(d) 2D inverse plug, and (e) uniform shear state for dense cases.
				The densities are $\bar\rho=0.156$, $0.305$, $0.723$, $0.780$ and $0.899$, respectively.
				The shear rate is fixed at ${\dot\gamma}^\ast=0.1$.}
	\label{fig:clusters}
\end{figure}

Next, we consider various density cases: 
$\bar\rho=0.0463$, $0.156$, $0.305$, $0.723$, $0.780$ and $0.899$.
Here, these values correspond to 
$L=60\sigma$, $40\sigma$, $32\sigma$, $24\sigma$, $23.4\sigma$ and $22.32\sigma$, respectively.
We set $T_0 = 1.4\varepsilon$ as the initial temperature, 
which is higher than the critical temperature for all densities,
and set ${\dot\gamma}^\ast=0.1$ as the shear rate.
Typical shapes of clusters for various densities are drawn in Fig. \ref{fig:evol}(b), \ref{fig:clusters}(a)--(d).
We found various shapes: a droplet (Fig. \ref{fig:evol}(b)), 2D plug (Fig. \ref{fig:clusters}(a)), 
2D plate (Fig. \ref{fig:clusters}(b)), 2D inverse plate (Fig. \ref{fig:clusters}(c)) and
2D inverse plug (Fig. \ref{fig:clusters}(d)).
These shapes for dilute densities are similar to those obtained 
by the previous work for macroscopic granular particles \cite{Conway}.
For denser region, the role of particles and vacancies are exchanged in the dilute region,
that is, the system has a particle hole symmetry with respect to the average density.
For example, 2D plug (Fig. \ref{fig:clusters}(a)) and 2D inverse plug (Fig. \ref{fig:clusters}(d)) are symmetric,
and this is also true for 2D plate (Fig. \ref{fig:clusters}(b)) and 2D inverse plate (Fig. \ref{fig:clusters}(c)),
however, we can not observe any inverse droplet in our simulations.
In addition, for liquid-solid coexistence region, there appear no clusters (Fig. \ref{fig:clusters}(e)).

\begin{figure}
	\begin{minipage}{0.49\hsize}
	\begin{overpic}[width=38mm]{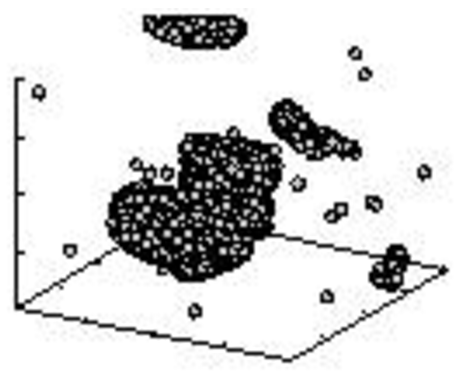}
		\put(31,-3){$x^\ast$}
		\put(-8,38){$y^\ast$}
		\put(97,5){$z^\ast$}
		\put(50,-6){(a)}
	\end{overpic}
	\end{minipage}
	\begin{minipage}{0.49\hsize}
	\begin{overpic}[width=38mm]{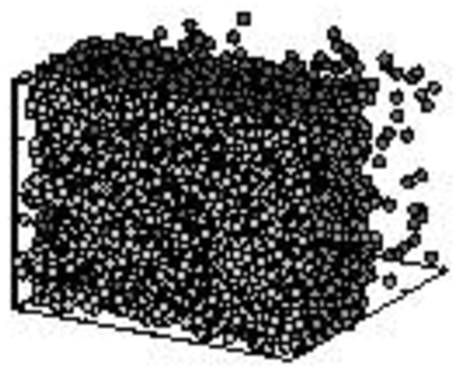}
		\put(31,-3){$x^\ast$}
		\put(-8,38){$y^\ast$}
		\put(97,5){$z^\ast$}
		\put(50,-6){(b)}
	\end{overpic}
	\end{minipage}
	\caption{Clusters for strong dissipation rates:
				(a) a cluster, whose shape is heart-like
				for $\bar\rho=0.0463$ ($\dot\gamma^\ast=0.1$, $\zeta^\ast = 3.0$)
				and (b) a cluster vertical to $z$-axis
				for $\bar\rho=0.305$ ($\dot\gamma^\ast=0.1$, $\zeta^\ast=0.1$).
				In both cases, the dissipation rate satisfies $\zeta \gg \zeta_{\rm cr}$.}
	\label{fig:other}
\end{figure}

For strong dissipation rate $\zeta \gg \zeta_{\rm cr}$,
the shapes of clusters depend on the initial conditions.
Typical shapes are plotted in Fig. \ref{fig:other}.
For instance, 
the shape of a droplet is not spherical but heart-like in the dilute case (Fig. \ref{fig:other}(a)),
but the shape strongly depends on the initial condition.
Indeed, we observe a spherical droplet similar to Fig. \ref{fig:evol}(b) 
even when we start from the identical shear rate and the density with a different initial configuration.
We can observe a cluster vertical to the $z$-axis in an intermediate density (Fig. \ref{fig:other}(b)).

Time evolution of $T_{\rm g}$ is similar to Fig. \ref{fig:temperature}.
The critical temperatures are approximately 
$T_{\rm cr}=1.26\varepsilon$, $1.28\varepsilon$ and $1.16\varepsilon$
for $\bar\rho=0.0463$, $0.305$ and $0.456$, respectively.
The former two temperatures are slightly higher than the equilibrium values \cite{Hansen, Smit},
while the latter is slightly lower than the equilibrium one.

\section{Conclusion}
\quad We studied the relationships between the shear rate and the dissipation rate in the system
for cohesive granular particles under a plane shear.
For the dilute case, we found that
the granular temperature decreases rapidly to zero after reaching the critical temperature $T_{\rm cr}$,
at which this characteristic time $t_{\rm cl}$ depends on the dissipation rate $\zeta$ as 
$t_{\rm cl}=\alpha(\zeta-\zeta_{\rm cr})^{-\beta}$, and the exponent $\beta$ is approximately given $0.8$
for $\bar\rho=0.0463$.

Depending on the average density, we observed various types of clusters.
We also found that the shapes of clusters for dilute cases
and those of inverse clusters for dense cases are almost symmetric with respect to the average density of the system.
Some shapes of clusters depend on the initial configuration for strongly dissipative cases.

\begin{theacknowledgments}
\quad Numerical computation in this work was carried out 
at the Yukawa Institute Computer Facility.
This work is partially supported by the Grant-in-Aid for the Global COE program
gThe Next Generation of Physics, Spun from Universality and Emergence from MEXT, Japan.''
\end{theacknowledgments}



\bibliographystyle{aipproc}   


\IfFileExists{\jobname.bbl}{}
 {\typeout{}
  \typeout{******************************************}
  \typeout{** Please run "bibtex \jobname" to optain}
  \typeout{** the bibliography and then re-run LaTeX}
  \typeout{** twice to fix the references!}
  \typeout{******************************************}
  \typeout{}
 }


\end{document}